\documentclass[conference]{IEEEtran}
\IEEEoverridecommandlockouts
\usepackage{cite}
\usepackage{amsmath,amssymb,amsfonts}
\usepackage{algorithmic}
\usepackage{graphicx}
\usepackage{textcomp}
\usepackage{url}
\usepackage{standalone}
\usepackage{adjustbox}
\usepackage{booktabs}
\usepackage{pifont}
\newcommand{\cmark}{\ding{51}}%
\newcommand{\xmark}{\ding{55}}%

\def\BibTeX{{\rm B\kern-.05em{\sc i\kern-.025em b}\kern-.08em
    T\kern-.1667em\lower.7ex\hbox{E}\kern-.125emX}}
\begin{document}

\title{
End-to-End Polyphonic Sound Event Detection Using Convolutional Recurrent Neural Networks with Learned Time-Frequency Representation Input
\thanks{The research leading to these results has received funding from the European Research Council under the European Union’s H2020 Framework Programme through ERC Grant Agreement 637422 EVERYSOUND. The authors wish to acknowledge CSC IT Center for Science, Finland, for providing computational resources.}
}

\author{\IEEEauthorblockN{Emre \c{C}ak\i{}r}
\IEEEauthorblockA{\textit{Tampere University of Technology, Finland} \\
emre.cakir@tut.fi}
\and
\IEEEauthorblockN{Tuomas Virtanen}
\IEEEauthorblockA{\textit{Tampere University of Technology, Finland} \\
tuomas.virtanen@tut.fi}
}

\maketitle

\begin{abstract}
Sound event detection systems typically consist of two stages: extracting hand-crafted features from the raw audio waveform, and learning a mapping between these features and the target sound events using a classifier. Recently, the focus of sound event detection research has been mostly shifted to the latter stage using standard features such as mel spectrogram as the input for classifiers such as deep neural networks. In this work, we utilize end-to-end approach and propose to combine these two stages in a single deep neural network classifier. The feature extraction over the raw waveform is conducted by a feedforward layer block, whose parameters are initialized to extract the time-frequency representations. The feature extraction parameters are updated during training, resulting with a representation that is optimized for the specific task. This feature extraction block is followed by (and jointly trained with) a convolutional recurrent network, which has recently given state-of-the-art results in many sound recognition tasks. The proposed system does not outperform a convolutional recurrent network with fixed hand-crafted features. The final magnitude spectrum characteristics of the feature extraction block parameters indicate that the most relevant information for the given task is contained in 0 - 3 kHz frequency range, and this is also supported by the empirical results on the SED performance.
\end{abstract}


\section{Introduction}

Sound event detection (SED) deals with the automatic identification of the sound events, \textit{i.e.}, sound segments that can be labeled as a distinctive concept in an audio signal. The aim of SED is to detect the onset and offset times for each sound event in an audio recording and associate a label with each of these events. At any given time instance, there can be either a single or multiple sound events present in the sound signal. The task of detecting a single event at a given time is called monophonic SED, and the task of detecting multiple sound events is called polyphonic SED. In recent years, SED has been proposed and utilized in various application areas including audio surveillance~\cite{foggia2015reliable}, urban sound analysis~\cite{bello2018sound}, multimedia event detection~\cite{wang2016audio} and smart home devices~\cite{krstulovic2018audio}.

SED has traditionally been approached as a two-stage problem: first, a time-frequency representation of the raw audio signal is extracted, then a classifier is used to learn the mapping between this representation and the target sound events. For the first stage, magnitude spectrograms, and human perception based methods such as mel spectrograms and mel frequency cepstral coefficients (MFCC) have been the most popular choices among SED researchers, and they have been used in a great portion of the submissions for the two recent SED challenges~\cite{mesaros2018detection,DCASE2017challenge}. For the second stage, deep learning methods such as convolutional and recurrent neural networks have recently been dominating the field with state-of-the-art performances~\cite{Lim2017,emre_TASLP2016,Adavanne2016}.

Using time-frequency representations are beneficial in the following ways. Compared to raw audio signal in time domain, frequency domain content matches better with the semantic information about sounds. In addition, the representation is 2-D, which makes the vast research on classifiers on image-based recognition tasks applicable to SED. Also, they are often more robust to noisy environments than raw audio signals (as the noise and the target sources can occupy different regions in the frequency domain), and the obtained performance is often better than using the raw audio signals as input to the second stage. On the other hand, especially for human perception based representations, it can be argued that these representations utilize domain knowledge to discard some information from the data, which could have been otherwise useful given the optimal classifier method.

\begin{figure*}
\centering
\includegraphics[width=0.9\linewidth]{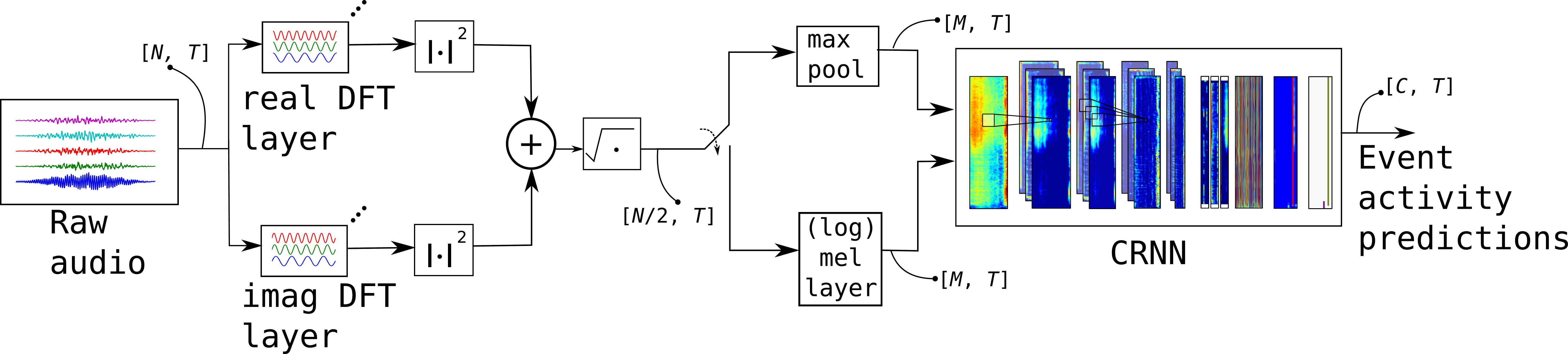}\\
\caption{Method framework. The method output shape in various stages of the framework is given in brackets.}\label{fig:framework}
\end{figure*}
\subsection{Related Work}
Recently, classifiers with high expression capabilities such as deep neural networks have been utilized to learn directly from raw representations in several areas of machine learning.
For instance, in image recognition, since the deep learning methods have been found to be highly effective with the works such as AlexNet~\cite{krizhevsky2012imagenet}, hand-crafted image features have been mostly replaced with raw pixel values as the inputs for the classifiers. 
For speech recognition, similar performance have been obtained for raw audio and log mel spectrograms in using convolutional, long-short term memory deep neural network (CLDNN) classifiers~\cite{sainath2015learning}. For music genre recognition, raw audio input for a CNN gives close performance to mel spectrograms~\cite{dieleman2014end}. Both~\cite{sainath2015learning} and~\cite{dieleman2014end} claim that when the magnitude spectra of the filter weights of the first convolutional layers are calculated and then visualized with the order of lowest dominant frequency to the highest, the resulting scale resembles the perception based scales such as mel and gammatone. For speech emotion recognition, a CLDNN classifier similar to~\cite{sainath2015learning} with raw audio input outperforms the standard hand-crafted features in the field, and it provides on-par performance with the state-of-the-art on a baseline dataset~\cite{trigeorgis2016adieu}.

However, when it comes to SED, hand-crafted time-frequency representations are still found to be more effective than raw audio signals as the classifier input. In~\cite{hertel2016comparing}, raw audio input performs considerably worse than concatenated magnitude and phase spectrogram features. In~\cite{Hou2017}, deep gated recurrent unit~(GRU) classifier with raw audio input ranks poorly compared to time-frequency representation based methods in DCASE2017 challenge sub-task on real-life SED~\cite{DCASE2017challenge}. Most likely due to the poor performance, the research on the end-to-end methods for SED has recently been very limited, and only two out of 200 submissions have used raw audio as classifier input (with low success) in DCASE2017 SED challenge~\cite{DCASE2017challenge}. As an attempt to move towards lower level input representations for SED, in~\cite{cakir2016filterbank}, magnitude spectrogram has been used an input to a deep neural network whose first layer weights were initialized with mel filterbank coefficients. 



\subsection{Contributions of this work}
In this work, we propose to use convolutional recurrent neural networks (CRNN) with learned time-frequency representation inputs for end-to-end SED. The most common time-frequency representations consist of applying some vector multiplications and basic math operations (such as sum, square and log) over raw audio signals divided into short time frames. This can be implemented in the form of a neural network layer, and the benefit is that the parameters used in the vector multiplications can be updated during network training to optimize the classifier performance for the given SED task. In this work, we investigate this approach by implementing magnitude spectrogram and (log) mel spectrogram extraction in the form of a feature extraction layer block, whose parameters can be adapted to produce an optimized time-frequency representation for the given task.
We then compare the adapted parameters with the initial parameters to gain insight on the neural network optimization process for the feature extraction block. To our knowledge, this is the first work to integrate and utilize domain knowledge into a deep neural network classifier in order to conduct end-to-end SED. 
The main differences between this work and the authors' earlier work on filterbank learning~\cite{cakir2016filterbank} are the input representation (raw audio vs. magnitude spectrogram), spectral domain feature extraction block using neural network layers and the classifier (CRNN vs. FNN and CNN).

\section{Method}

The input $\mathbf{X} \in \mathbb{R}^{N \times T}$ consists of $T$ frames of raw audio waveforms sampled of $N$ samples with sampling rate $F_s$, and Hamming window with $N$ samples is applied to each frame. Initially (\textit{i.e.} before the network training), the output of the feature extraction block is either max pooled magnitude spectrogram, mel spectrogram or log mel spectrogram. The method framework is illustrated in Figure~\ref{fig:framework}. 

\subsection{Feature Extraction block}

The input $\mathbf{X}$ to the feature extraction block is fed through two parallel feedforward layers, $l^\mathrm{re}$ and $l^\mathrm{im}$, each with $N$ neurons with linear activation function and no bias. The weights of these two layers, namely $\mathbf{W}^\mathrm{re} \in \mathbb{R}^{\frac{N}{2} \times N}$ and $\mathbf{W}^\mathrm{im} \in \mathbb{R}^{\frac{N}{2} \times N}$, are initialized so that the outputs of these layers for each frame $\mathbf{X}_{:,t}$ ($t=1,...T$) would correspond to the real and the imaginary parts of the 
discrete Fourier transform (DFT):
\begin{equation}
\begin{aligned}
\mathbf{F}_{k, t} &= \sum \limits_{n=0}^{N-1}\mathbf{X}_{n,t}[\cos(2\pi kn/N) -i\cdot \sin(2\pi kn/N)] \\
\mathbf{W}^\mathrm{re}_{k, n} &\leftarrow \cos(2\pi kn/N) \\
\mathbf{W}^\mathrm{im}_{k, n} &\leftarrow \sin(2\pi kn/N) \\
\mathbf{Z}^\mathrm{re}_{k, t} &= \sum \limits_{n=0}^{N-1}\mathbf{W}^\mathrm{re}_{k, n}\mathbf{X}_{n,t} \\
\mathbf{Z}^\mathrm{im}_{k, t} &= \sum \limits_{n=0}^{N-1}\mathbf{W}^\mathrm{im}_{k, n}\mathbf{X}_{n,t}
\end{aligned}
\end{equation}
for $k=0,1,...,\frac{N}{2}-1$ and $n=0,...,N-1$, where $\mathbf{Z}$ is the weighted output for each feedforward layer. The reason for taking only the first half of the DFT bins is that the raw audio waveform input $\mathbf{X}$ is purely real, resulting with a symmetric magnitude spectrum. Each weight vector $\mathbf{W}_{k, :}$ can be deemed as an individual sinusoidal filter. For both $l^\mathrm{re}$ and $l^\mathrm{im}$, the outputs given the input $\mathbf{X}$ is calculated using the same weights $\mathbf{W}^\mathrm{re}$ and $\mathbf{W}^\mathrm{im}$ for each of the $T$ frames. Both layers are followed by a square operation, the outputs of the layers are summed, and finally a square root operator results with the magnitude spectrogram $\mathbf{S} \in \mathbb{R}^{\frac{N}{2} \times T}$:
\begin{equation}
\mathbf{S}_{k, t} = |\mathbf{F}_{k, t}| = \sqrt[]{({\mathbf{Z}^\mathrm{re}_{k, t}})^2 + ({\mathbf{Z}^\mathrm{im}_{k, t}})^2}
\end{equation}
At this stage, $\mathbf{S}$ can be directly fed as input to a CRNN classifier, or it can be further processed to obtain $M$ (log) mel spectrogram using a feedforward layer with $M$ neurons, rectified linear unit (ReLU) activations and no bias:
\begin{equation}
\mathbf{Z}^\mathrm{mel}_{m, t} = \max(0, \sum \limits_{k=0}^{N/2-1}\mathbf{W}^\mathrm{mel}_{m,k}\mathbf{S}_{k,t}) 
\end{equation}
for $m=0,1,...M-1$. The weights $\mathbf{W}^\mathrm{mel}$ of this layer is initialized with the mel filterbank coefficients in the similar manner with~\cite{cakir2016filterbank} and log compression is used in part of the experiments as
\begin{equation}
\mathbf{Z}^\mathrm{logmel} = \log(\mathbf{Z}^\mathrm{mel} + \epsilon)
\end{equation}
where $\epsilon=0.001$ is used to avoid numerical errors. The parameters $\mathbf{W}^\mathrm{mel}$ are obtained from \textit{Librosa}~\cite{librosa} package and the center frequencies for each mel band are calculated using O'Shaughnessy's formula~\cite{o1987speech}. For the experiments where this layer is utilized, the weights $\mathbf{W}^\mathrm{re}$ and $\mathbf{W}^\mathrm{im}$ are kept fixed, as explained in Table~\ref{tab:learned}.

In our experiments while using $\mathbf{S}$ directly as the input for CRNN, we observed that when the number of features for $\mathbf{S}$ is dropped from $\frac{N}{2}$ to $M$ by using max-pooling in frequency domain, the computation time is substantially reduced with very limited decrease in accuracy. Hence, we followed this approach when the mel feature layer is omitted.

\subsection{Convolutional Recurrent block}
\label{sec:crnn}
Following the same approach with~\cite{emre_TASLP2016}, the CRNN block consists of three parts: 
\begin{enumerate}
	\item{convolutional layers with ReLU activations and non-overlapping pooling over frequency axis}
    \item{gated recurrent unit (GRU)~\cite{cho2014properties} layers, and}
    \item{a single feedforward layer with $C$ units and sigmoid activation, where $C$ is the number of target event classes.}
\end{enumerate}

The output of the feature extraction block, \textit{i.e.}, a sequence of feature vectors, is fed to the convolutional layers and the activations from the filters of the last convolutional layer are stacked over the frequency axis and fed to the GRU layers. For each frame, GRU layer activations are calculated using both the current frame input and the previous frame outputs. Finally, the GRU layer activations are fed to the fully-connected layer. The output of this final layer is treated as the event activity probability for each event. The aim of the network learning is to get the estimated frame-level class-wise event activity probabilities as close as to their binary target outputs, where target output is 1 if an event class is present in a given frame, and 0 vice versa. In the usage case, the estimated frame-level event activity probabilities are thresholded with 0.5 to obtain binary event activity predictions. More detailed explanation about CRNN block can be found in~\cite{emre_TASLP2016}.

The network is trained with back-propagation through time using Adam optimizer~\cite{adamKeras} with learning rate $10^{-3}$, binary cross-entropy as the loss function and for maximum 300 epochs. In order to reduce overfitting of the model, early stopping was used to stop training if the validation data frame-level F1 score did not improve for 65 epochs. For regularization, batch normalization~\cite{batchNorm} was employed in convolutional layers and dropout~\cite{Dropout} with rate 0.25 was employed in convolutional and recurrent layers. Keras deep learning library~\cite{chollet2015keras} was used to implement the network.


\begin{table}[!t]
\centering
\caption{A table  showing which weight matrices are learned for each experiment. ~\cmark ~stands for learned, ~\xmark ~stands for fixed, and - stands for not utilized in the experiment.}
\label{tab:learned} \normalsize
\resizebox{0.8\linewidth}{!}{%

\begin{tabular}{*{4}{c}}
\toprule
Learned? & $\mathbf{W}^\mathrm{re}$ & $\mathbf{W}^\mathrm{im}$ & $\mathbf{W}^\mathrm{mel}$ \\
\midrule
DFT learned & \cmark & \cmark & - \\ 
Mel learned & \xmark & \xmark & \cmark \\
Log mel learned & \xmark & \xmark & \cmark \\
\bottomrule
\end{tabular}
}
\end{table}

\section{Evaluation}

\subsection{Dataset}
The dataset used in this work is called \textit{TUT-SED Synthetic 2016}. It is a publicly available polyphonic SED dataset, which consists of synthetic mixtures created by mixing isolated sound events from 16 sound event classes. Polyphonic mixtures were created by mixing 994 sound event samples with the sampling rate 44.1 kHz. From the 100 mixtures created, 60\% are used for training, 20\% for testing and 20\% for validation. The total length of the data is 566 minutes. Different instances of the sound events are used to synthesize the training, validation and test partitions. Mixtures were created by randomly selecting event instance and from it, randomly, a segment of length 3-15 seconds. Mixtures do not contain any additional background noise. Dataset creation procedure explanation and metadata can be found in the web page~\footnote{http://www.cs.tut.fi/sgn/arg/taslp2017-crnn-sed/tut-sed-synthetic-2016} hosting the dataset.

\subsection{Evaluation Metrics and Experimental Setup}
The evaluation metrics used in this work are frame-level F1 score and error rate. F1 score is the harmonic mean of precision and recall, and error rate is the sum of the rate of insertions, substitutions and deletions. Both metrics are calculated in the same manner with~\cite{emre_TASLP2016} and they are explained in more detail in~\cite{mesaros2016metrics}.

The input $\mathbf{X}$ to the feature extraction block consists of a sequence of 40 ms length frames with 50\% overlap. The number of frames in the sequence is $T=256$  which corresponds to 2.56 seconds of raw audio. The audio signals have been resampled from the original rate of 44.1 kHz to 8, 16 and 24 kHz in different experiments, which corresponds to $N=$~160, 320, and 480 features for each frame, respectively. 
This is done both to investigate the effect of discarding the information from higher frequencies, and also to reduce the memory requirements to be able to run experiments with a decent sized network and batch size. At the max pooling (or mel) layer of the feature extraction block, the number of features is set to $M=$~40.

\begin{table}[!t]
\centering
\caption{Frame-level F1 score $F1_{\textnormal{frm}}$ and error rate $ER_{\textnormal{frm}}$ results for different time-frequency representation methods and sampling rates. 
"DFT" stands for magnitude spectrogram using linear frequency scale, "Mel" stands for mel spectrogram, "fixed" and "learned" stands for whether the weights of the feature extraction block are kept fixed or updated during training.  
}
\label{tab:results} \normalsize
\resizebox{\linewidth}{!}{%
\begin{tabular}{ @{} l *{1}{c c }} 
\toprule 

Method & $F1_{\textnormal{frm}}$ & $ER_{\textnormal{frm}}$ \\
\midrule

DFT 8 kHz fixed 					& 60.8$\pm$0.8 &0.55$\pm$0.01 \\
DFT 8 kHz learned 					& 60.8$\pm$0.8 &0.55$\pm$0.01 \\
\midrule
Mel 8 kHz fixed 					& 60.8$\pm$0.9 &0.55$\pm$0.01 \\
Mel 8 kHz learned 					& 61.0$\pm$0.8 &0.56$\pm$0.01 \\
\midrule
Log mel 8 kHz fixed 					& 63.1$\pm$0.6 &0.52$\pm$0.01 \\
Log mel 8 kHz learned 					& 58.6$\pm$1.6 &0.56$\pm$0.01 \\
\midrule
DFT 16 kHz fixed 					& 61.9$\pm$0.9 &0.54$\pm$0.01 \\
DFT 16 kHz learned 					& 60.1$\pm$1.7 &0.58$\pm$0.03 \\
\midrule
Mel 16 kHz fixed 					& 62.3$\pm$0.7 &0.54$\pm$0.01 \\
Mel 16 kHz learned 					& 60.6$\pm$0.9 &0.57$\pm$0.02 \\
\midrule
Log mel 16 kHz fixed 					& 65.8$\pm$1.4 &0.50$\pm$0.01 \\
Log mel 16 kHz learned 					& 59.9$\pm$1.3 &0.56$\pm$0.01 \\
\midrule
DFT 24 kHz learned 					& 58.1$\pm$1.6 &0.59$\pm$0.03 \\
\midrule

Log mel 44.1 kHz fixed~\cite{emre_TASLP2016} 					& 66.4$\pm$0.6 &0.48$\pm$0.01 \\

\bottomrule
\end{tabular}
}
\end{table} 

In order to find the optimal network hyper-parameters, a grid search was performed, and the hyper-parameter set resulting with the best frame-level F1 score on the validation data was used in the evaluation. The grid search consists of every possible combination of the following hyper-parameters: the number of convolutional filters / recurrent hidden units (the same amount for both) \{96, 256\}; the number of recurrent layers \{1, 2, 3\}; and the number of convolutional layers \{1, 2, 3 ,4\} with the following frequency max pooling arrangements after each convolutional layer \{(4), (2, 2), (4, 2), (8, 5), (2, 2, 2), (5, 4, 2), (2, 2, 2, 1), (5, 2, 2, 2)\}. Here, the numbers denote the number of features at each max pooling step; e.g., the configuration (5, 4, 2) pools the original 40 features in a single feature in three stages: 40$\rightarrow$8$\rightarrow$2$\rightarrow$1. This grid search process is repeated for every experiment setup in Table~\ref{tab:results} (except the last experiment, where a similar grid search has been performed earlier for that work).

After finding the optimal hyper-parameters, each experiment is run ten times with different random seeds to reflect the effect of random weight initialization in convolutional recurrent block of the proposed system. The mean and the standard deviation (given after $\pm$) of these experiments are provided.

\subsection{Results}

The effect of feature extraction with learned parameters have been investigated and compared with the fixed feature extraction parameters in Table~\ref{tab:results}. For both frame-level F1 score and error rate metrics, experiments with fixed feature extraction parameters often outperform the learned feature extraction methods in their corresponding sampling rates. In addition, the experiments with fixed parameters benefit from the increased sampling rate, whereas the performance does not improve for learned feature extraction parameters with higher sampling rates. One should also note that the F1 score using both learned and fixed parameters with 8 kHz sampling rate is 60.8\%. Although there is some drop in performance from the highest F1 score of 66.4\% at 44.1 kHz, it is still remarkable performance considering that about 82\% of the frequency domain content of the original raw audio signal is discarded in the resampling process from 44.1 kHz to 8 kHz. This emphasizes the importance of low frequency components for the given SED task. Since the computational load due to high amount of data in the raw audio representations is one of the concerns for end-to-end SED systems, it can be considered to apply a similar resampling process for the end-to-end SED methods in the future. 

\begin{figure}
  \begin{center}
  \includegraphics[width=3.5in, height=1.5in]{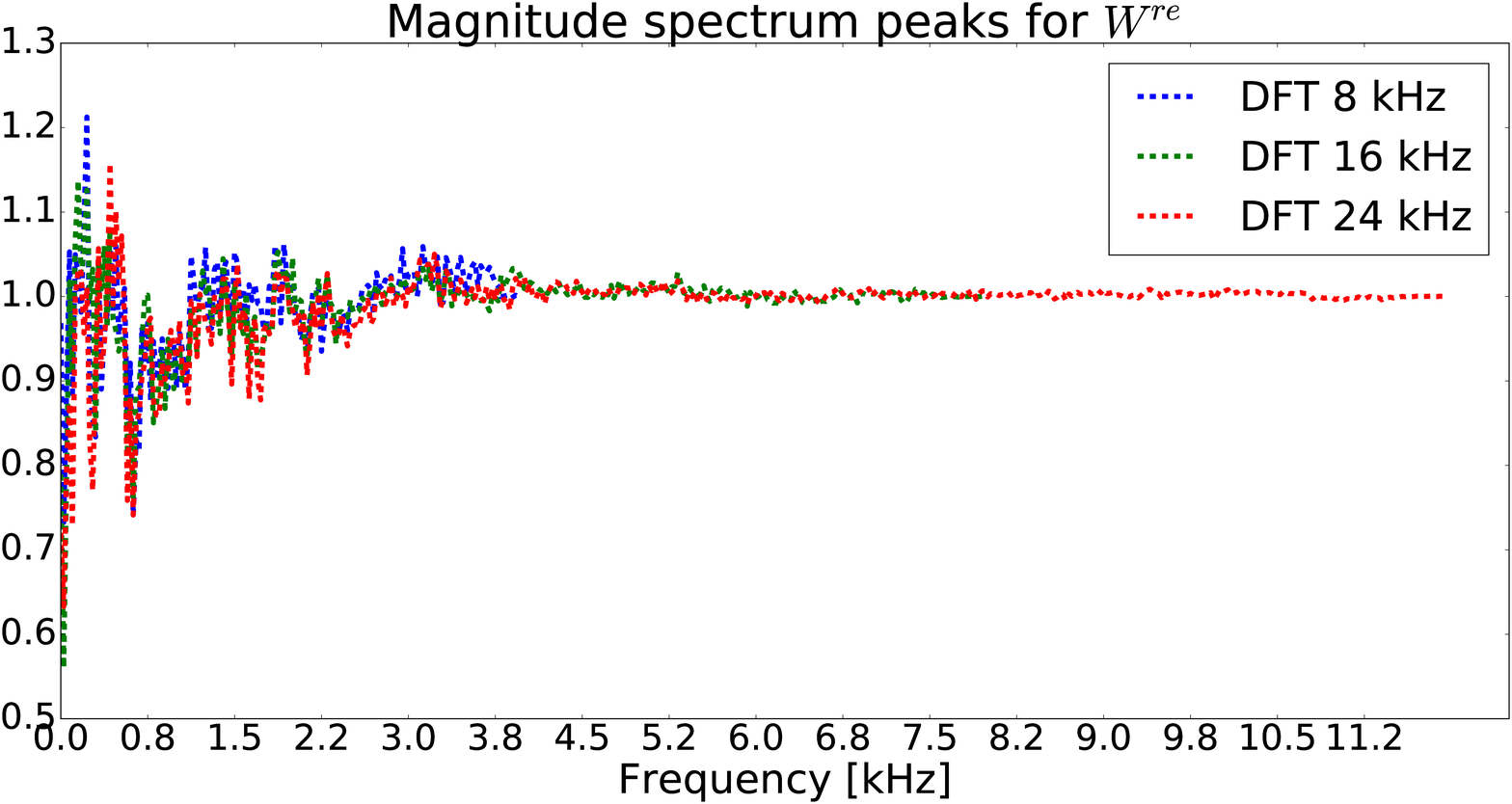}\\
  \includegraphics[width=3.5in, height=1.5in]{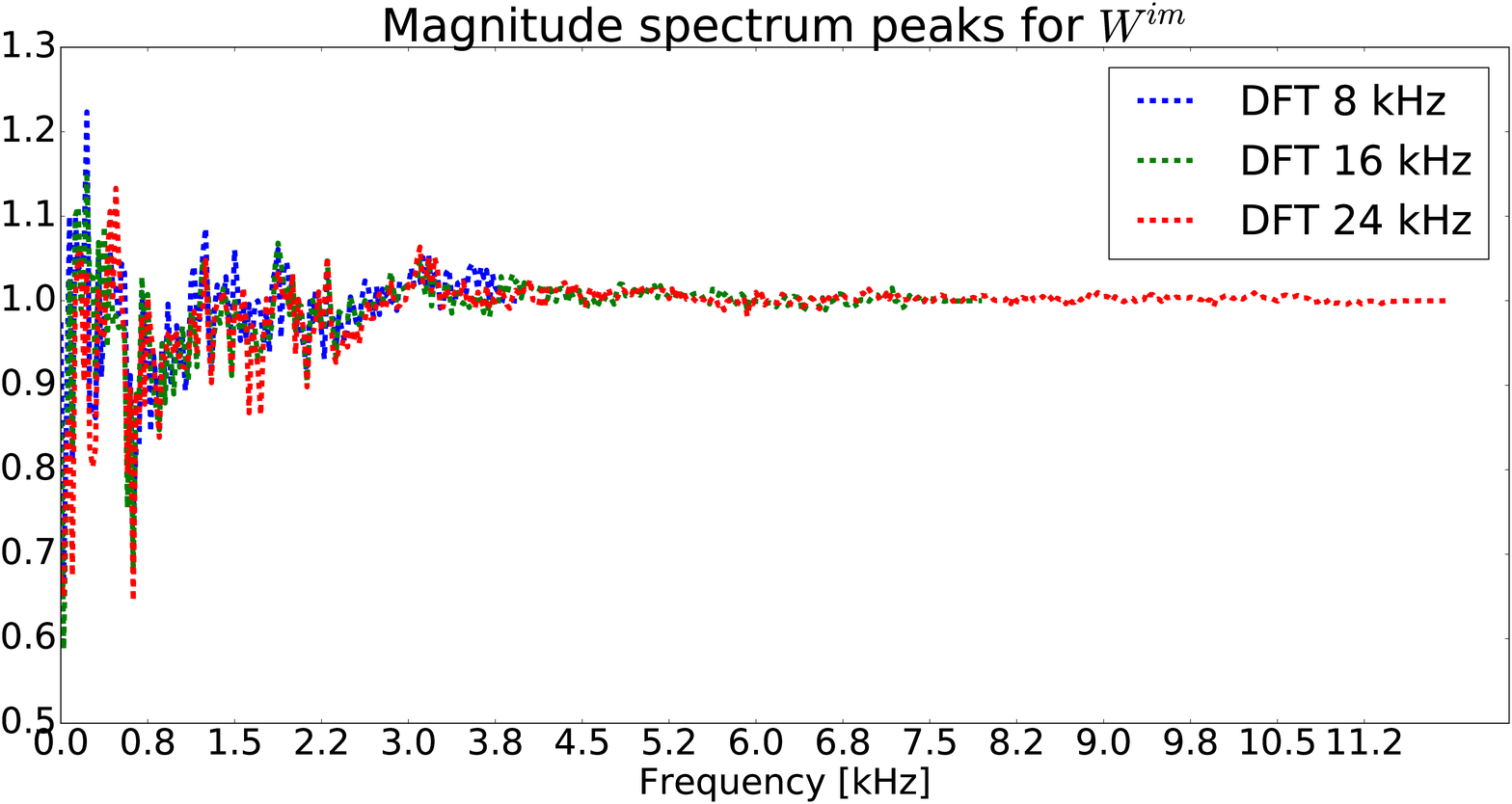}\\
  \includegraphics[width=3.5in, height=1.5in]{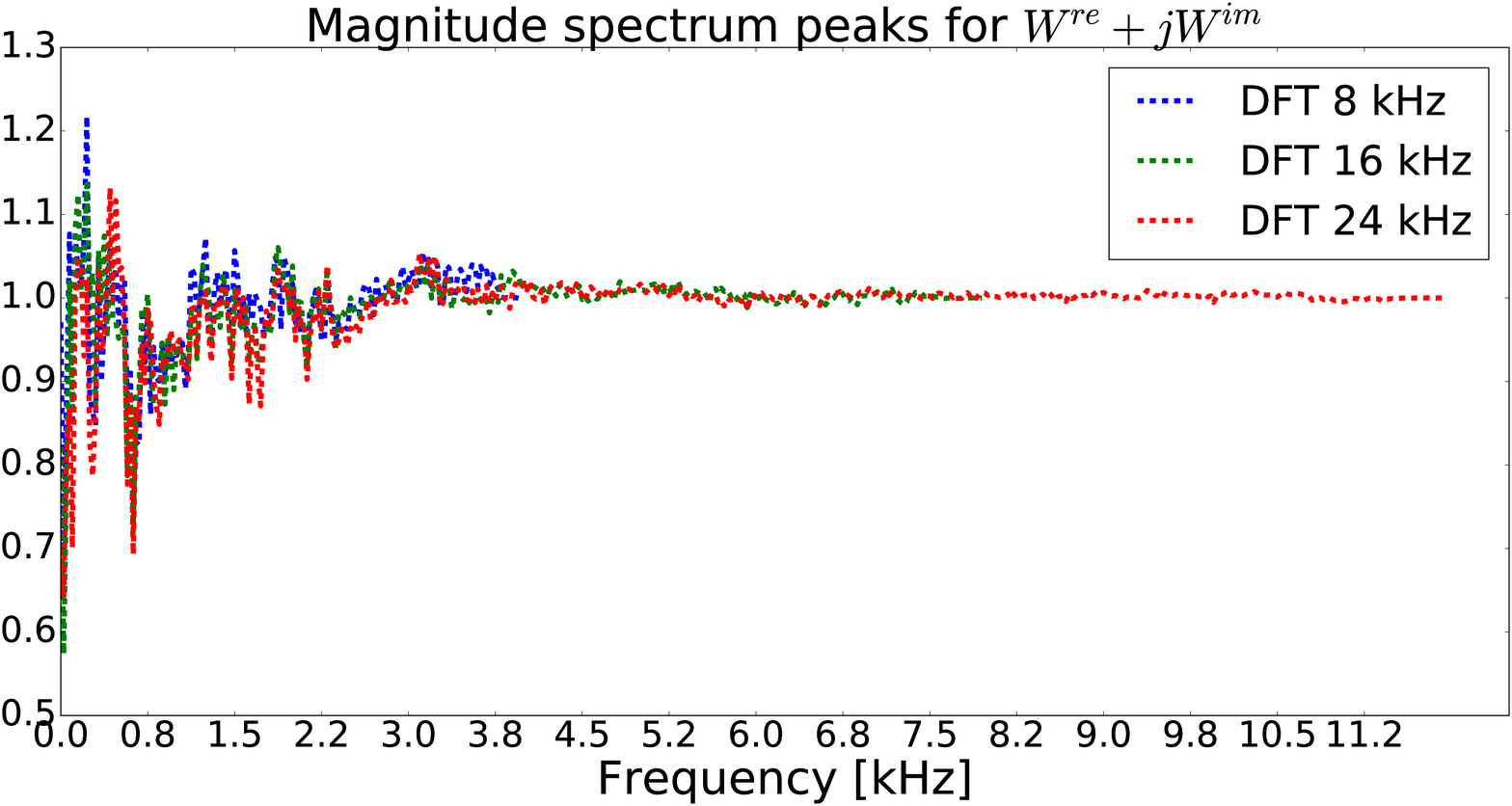}\\
  \includegraphics[width=3.62in, height=1.5in]{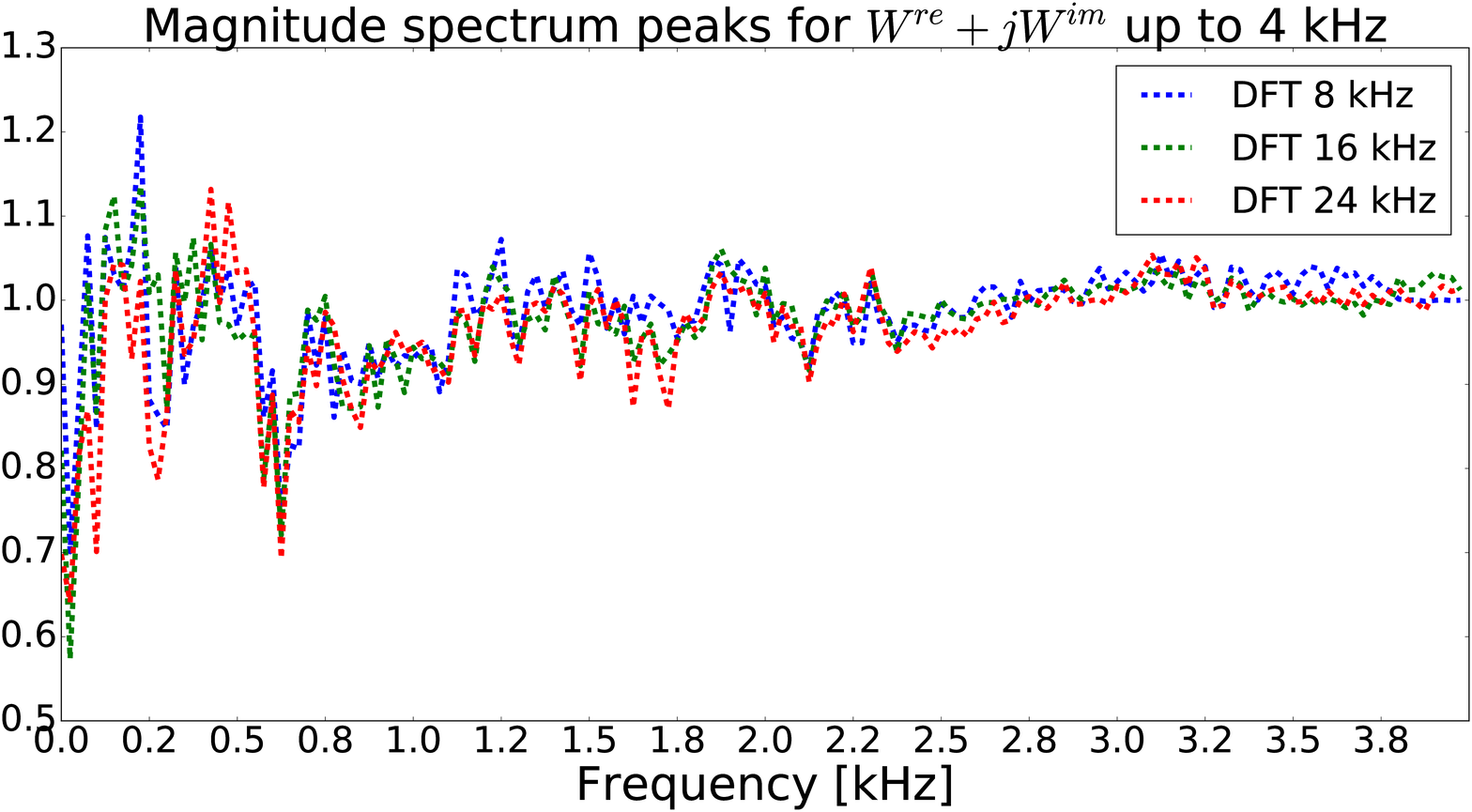}\\

\caption{Magnitude spectrum peaks for the real ($\mathbf{W}^\mathrm{re}$) and imaginary ($\mathbf{W}^\mathrm{im}$) DFT layer filters after the training. The amplitude of the peak for each filter is positioned at the center frequency of the corresponding filter, resulting with a line plot covering the whole frequency range for the experiment with given sampling rate.}\label{fig:mag_spec}
  \end{center}
\end{figure}
\begin{figure}[t]
  \begin{center}
  \includegraphics[width=3.58in, height=2in]{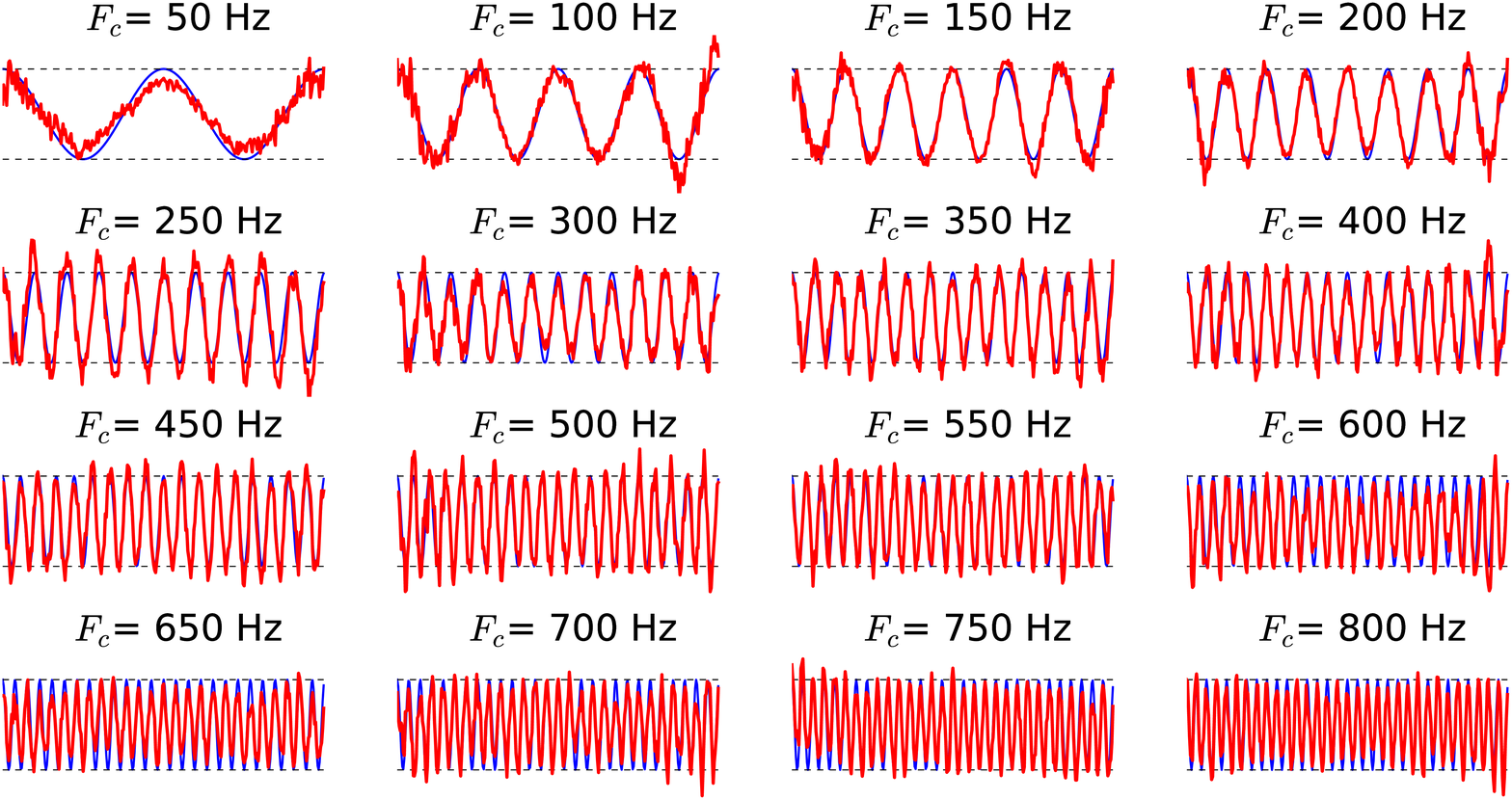}\\
\center (a)\\

\includegraphics[width=3.58in, height=2in]{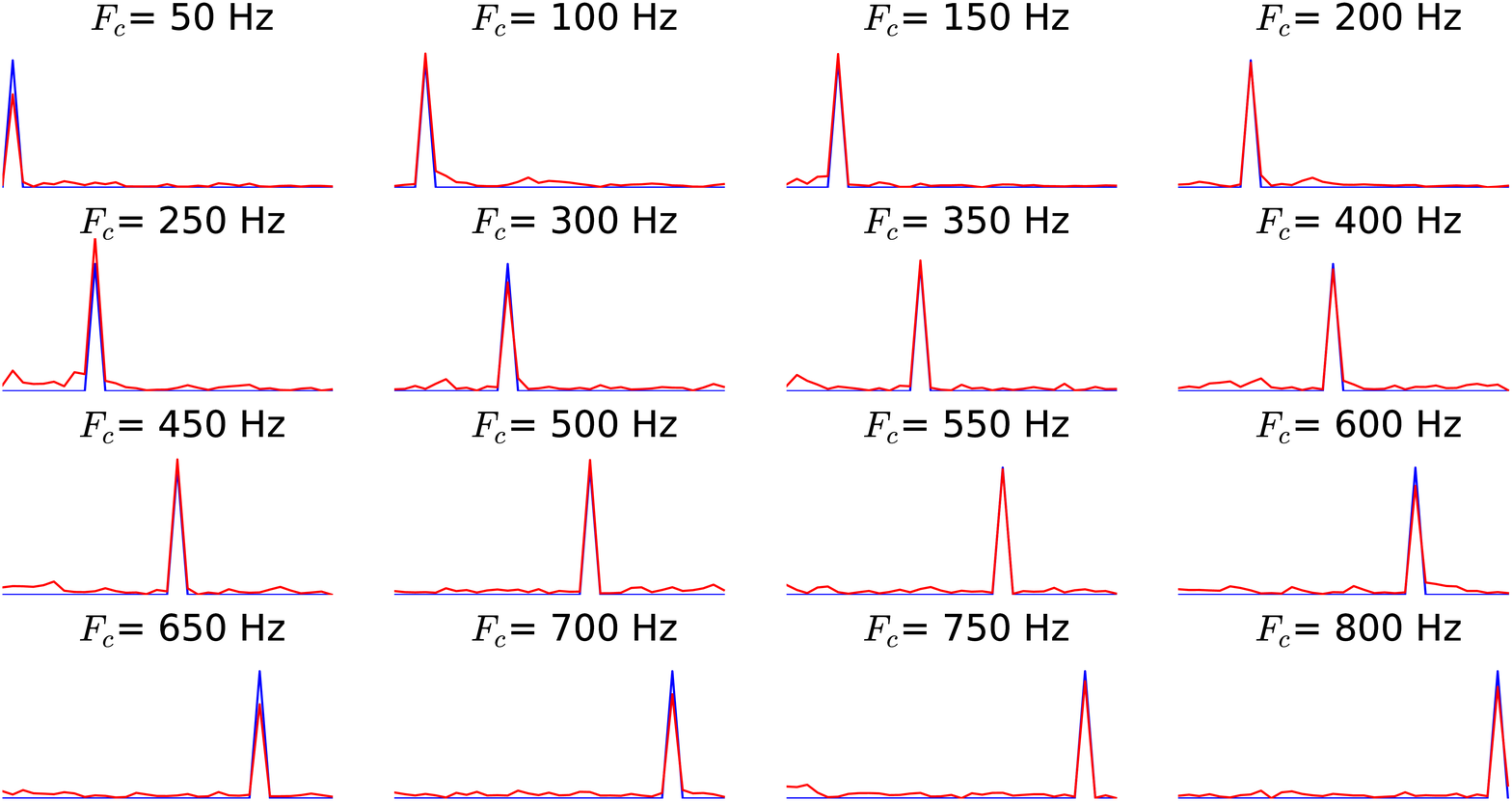}\\
\center (b)
\caption{(a): Real ($\mathbf{W}^\mathrm{re}$) DFT layer filters with different center frequencies $F_c$ and (b): their magnitude spectra. Blue plot represents initial values, and red plot represents the values after the training. Horizontal dashed lines at -1 and 1 mark the initial maximum and minimum values for the filters.}\label{fig:real_filts}
  \end{center}
\end{figure}

In order to investigate how the original parameters of the feature extraction block have been modified during the training, the magnitude spectrum peak, \textit{i.e.} the maximum value of the magnitude spectrum, of the trained weights for $\mathbf{W}^\mathrm{re}_k$, $\mathbf{W}^\mathrm{im}_k$, and $\mathbf{W}^\mathrm{re}_k + i\cdot\mathbf{W}^\mathrm{im}_k$ are calculated for each filter $k$. Without network training, these weights represent sinusoid signals, therefore the magnitude spectrum of each filter is equal to a single impulse at the center frequency of the filter, whose amplitude equals to the number of filters. At the end of the training, the peak of the magnitude spectrum for each filter stays at the center frequency of the filter, while the amplitude of the peak is either increased or decreased to a certain degree. In order to visualize the change in the peak amplitude, the peak amplitude positioned at the center frequency for each filter after training is given in Figure~\ref{fig:mag_spec}. 
The same analysis is repeated for different experiments using raw audio inputs with different sampling rates (8 kHz, 16 kHz and 24 kHz) as input to their feature extraction block which initially calculates the pooled magnitude spectrogram. The magnitude spectrum peaks for each experiment is scaled with the number of filters for visualization purposes, and therefore each peak is equal to 1 before the training. The three observations that can be made from Figure~\ref{fig:mag_spec} is
\begin{itemize}
\item Although each of these three systems have different CRNN architectures (grid search for each system results with different hyper-parameter set) and their raw audio input is sampled with different rates, the magnitude spectrum peaks possess very similar characteristics. For all three experiments, the peaks are modified the most for the frequencies below around 3 kHz, and there is little to no change in peak amplitudes after 4 kHz. This may indicate that the most relevant information for the given SED task is in 0-4 kHz region. Although the authors cannot conclude this, it is empirically supported to a certain degree with the results presented in Table~\ref{tab:results}. 
Even though the amount of data from the raw audio input sampled with 44.1 kHz is substantially reduced by resampling with 8 and 16 kHz, the performance drop is limited. 
\item For all three experiments, the change in the magnitude spectrum peaks is not monotonic in the frequency axis. Some of the peaks in the low frequency range are boosted, but there are also other peaks in the same frequency region that are significantly suppressed. This implies a different optimal time-frequency representation than both magnitude and mel spectrogram. One should also bear in mind that this learned representation is task-specific, and the same approach for other classification tasks may lead to a different ad-hoc magnitude spectrum peak distribution.
\item Although they represent different branches of the feature extraction block (and therefore are updated with different gradients), the magnitude spectrum peaks of $\mathbf{W}^\mathrm{re}$ and $\mathbf{W}^\mathrm{im}$ are modified in a very similar manner at the end of the training. 

\end{itemize}

The learned filters with the center frequencies up to 800 Hz with the sampling rate 8 kHz and their magnitude spectra have been visualized in Figure~\ref{fig:real_filts}. The neural network training process seemingly do not result with a shift in the center frequencies of the filter. On the other hand, it should be noted that in addition to the peak at the center frequency, the magnitude spectrum of each filter consists of other components with smaller amplitude values spread over the frequency range, which reflects that the pure sinusoid property of the filters are lost.

\begin{figure}[t]
  \begin{center}
  \includegraphics[width=3.58in, height=1in]{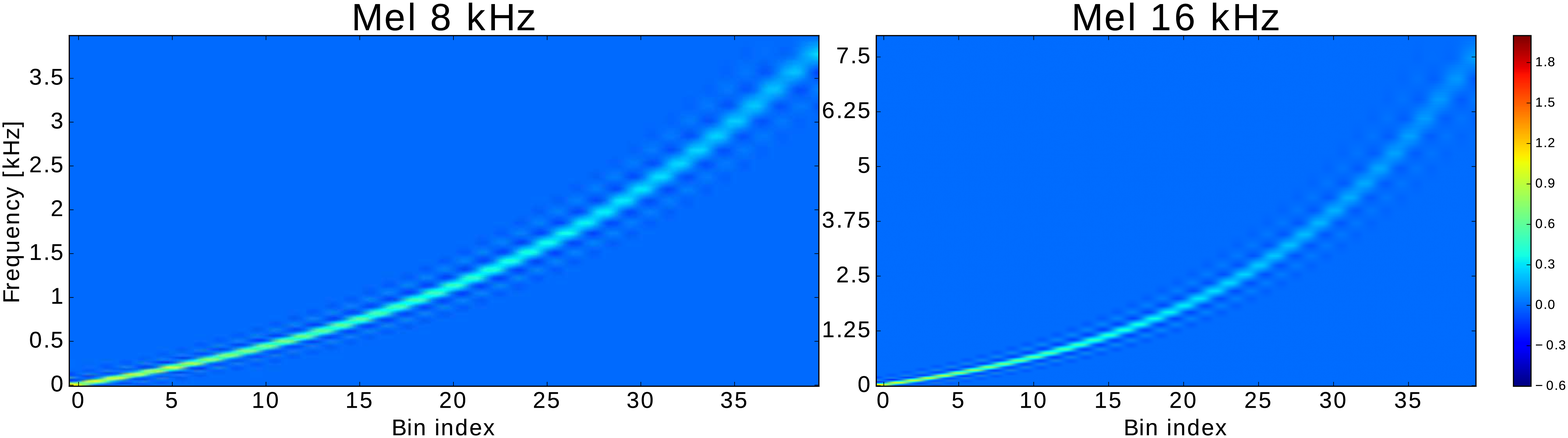}\\
\center (a)\\

\includegraphics[width=3.58in, height=1.8in]{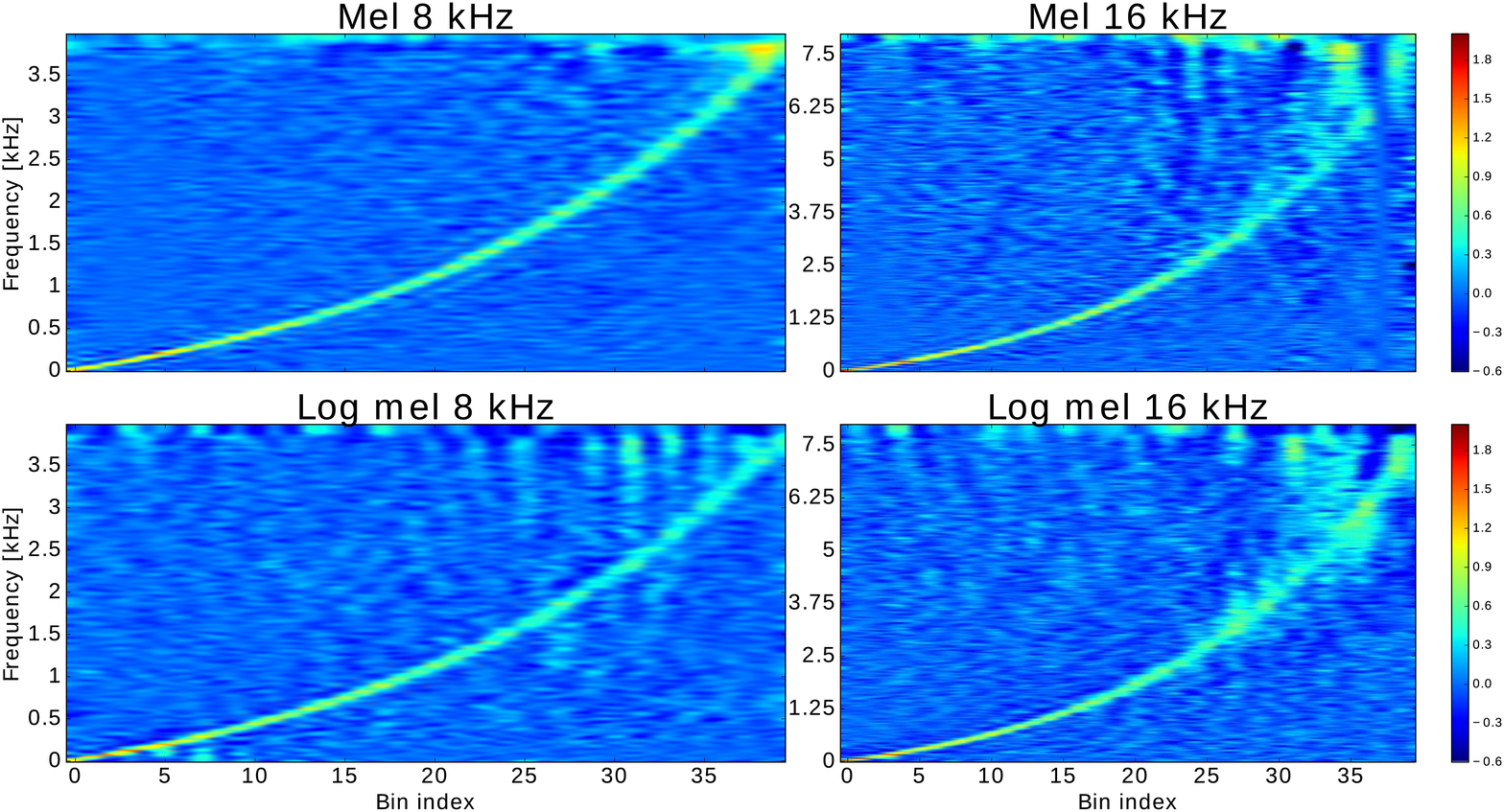}\\
\center (b)
\caption{(a): Initial, and (b): learned mel filterbank responses.}\label{fig:mel_filts}
  \end{center}
\end{figure}

For the experiments where the mel layer is utilized, the learned mel filterbank responses are visualized in Figure~\ref{fig:mel_filts}. One common point among the responses is that the filterbank parameters covering lower frequency range have been emphasized. The learned filterbank response that is the most resembling its initial response belongs to mel layer with 8 kHz sampling rate, which also performs the best among these four experiments with 61\% F1 score, as presented in Table~\ref{tab:results}. For the response of both mel and log mel layers with sampling rate 16 kHz, the parameters covering higher frequency range have been emphasized, and the filter bandwidths for higher frequencies have been increased. However this does not result with an improved performance, as these experiments provide 60.6\% and 59.9\% F1 score, respectively.

\section{Conclusion}
In this work, we propose to conduct end-to-end polyphonic SED using learned time-frequency representations as input to a CRNN classifier. The classifier is fed by a neural network layer block, whose parameters are initialized to extract common time-frequency representation methods over raw audio signals. These parameters are then updated through the training process for the given SED task. The performance of this method is slightly lower than directly using common time-frequency representations as input. During the network training, regardless of the input sampling rate and the neural network configuration, the magnitude response of the feature extraction block parameters have been significantly altered for the lower frequencies (below 4 kHz), and stayed mostly the same for higher frequencies.

\bibliographystyle{IEEEtran}
\bibliography{IEEEabrv,Bibliography}

\begin{thebibliography}{10}
\providecommand{\url}[1]{#1}
\csname url@rmstyle\endcsname
\providecommand{\newblock}{\relax}
\providecommand{\bibinfo}[2]{#2}
\providecommand\BIBentrySTDinterwordspacing{\spaceskip=0pt\relax}
\providecommand\BIBentryALTinterwordstretchfactor{4}
\providecommand\BIBentryALTinterwordspacing{\spaceskip=\fontdimen2\font plus
\BIBentryALTinterwordstretchfactor\fontdimen3\font minus
  \fontdimen4\font\relax}
\providecommand\BIBforeignlanguage[2]{{%
\expandafter\ifx\csname l@#1\endcsname\relax
\typeout{** WARNING: IEEEtran.bst: No hyphenation pattern has been}%
\typeout{** loaded for the language `#1'. Using the pattern for}%
\typeout{** the default language instead.}%
\else
\language=\csname l@#1\endcsname
\fi
#2}}

\bibitem{foggia2015reliable}
P.~Foggia, N.~Petkov, A.~Saggese, N.~Strisciuglio, and M.~Vento, ``Reliable
  detection of audio events in highly noisy environments,'' \emph{Pattern
  Recognition Letters}, vol.~65, pp. 22--28, 2015.

\bibitem{bello2018sound}
J.~P. Bello, C.~Mydlarz, and J.~Salamon, ``Sound analysis in smart cities,'' in
  \emph{Computational Analysis of Sound Scenes and Events}.\hskip 1em plus
  0.5em minus 0.4em\relax Springer, 2018, pp. 373--397.

\bibitem{wang2016audio}
Y.~Wang, L.~Neves, and F.~Metze, ``Audio-based multimedia event detection using
  deep recurrent neural networks,'' in \emph{2016 IEEE Int. Conf. on Acoustics,
  Speech and Signal Processing (ICASSP)}.\hskip 1em plus 0.5em minus
  0.4em\relax IEEE, 2016, pp. 2742--2746.

\bibitem{krstulovic2018audio}
S.~Krstulovi{\'c}, ``Audio event recognition in the smart home,'' in
  \emph{Computational Analysis of Sound Scenes and Events}.\hskip 1em plus
  0.5em minus 0.4em\relax Springer, 2018, pp. 335--371.

\bibitem{mesaros2018detection}
A.~Mesaros, T.~Heittola, E.~Benetos, P.~Foster, M.~Lagrange, T.~Virtanen, and
  M.~D. Plumbley, ``Detection and classification of acoustic scenes and events:
  Outcome of the {DCASE} 2016 challenge,'' \emph{IEEE/ACM Transactions on
  Audio, Speech, and Language Processing}, vol.~26, no.~2, pp. 379--393, 2018.

\bibitem{DCASE2017challenge}
A.~Mesaros, T.~Heittola, A.~Diment, B.~Elizalde, A.~Shah, E.~Vincent, B.~Raj,
  and T.~Virtanen, ``{DCASE} 2017 challenge setup: Tasks, datasets and baseline
  system,'' in \emph{Proceedings of the Detection and Classification of
  Acoustic Scenes and Events 2017 Workshop (DCASE2017)}, November 2017.

\bibitem{Lim2017}
H.~Lim, J.~Park, and Y.~Han, ``Rare sound event detection using {1D}
  convolutional recurrent neural networks,'' DCASE2017 Challenge, Tech. Rep.,
  September 2017.

\bibitem{emre_TASLP2016}
E.~Cakir, G.~Parascandolo, T.~Heittola, H.~Huttunen, and T.~Virtanen,
  ``Convolutional recurrent neural networks for polyphonic sound event
  detection,'' \emph{IEEE/ACM Transactions on Audio, Speech, and Language
  Processing}, vol.~25, no.~6, pp. 1291--1303, 2017.

\bibitem{Adavanne2016}
S.~Adavanne, G.~Parascandolo, P.~Pertil{\"a}, T.~Heittola, and T.~Virtanen,
  ``Sound event detection in multichannel audio using spatial and harmonic
  features,'' DCASE2016 Challenge, Tech. Rep., September 2016.

\bibitem{krizhevsky2012imagenet}
A.~Krizhevsky, I.~Sutskever, and G.~E. Hinton, ``Imagenet classification with
  deep convolutional neural networks,'' in \emph{Advances in neural information
  processing systems}, 2012, pp. 1097--1105.

\bibitem{sainath2015learning}
T.~N. Sainath, R.~J. Weiss, A.~Senior, K.~W. Wilson, and O.~Vinyals, ``Learning
  the speech front-end with raw waveform {CLDNN}s,'' in \emph{Proc.
  Interspeech}, 2015.

\bibitem{dieleman2014end}
S.~Dieleman and B.~Schrauwen, ``End-to-end learning for music audio,'' in
  \emph{IEEE International Conference on Acoustics, Speech and Signal
  Processing (ICASSP)}.\hskip 1em plus 0.5em minus 0.4em\relax IEEE, 2014, pp.
  6964--6968.

\bibitem{trigeorgis2016adieu}
G.~Trigeorgis, F.~Ringeval, R.~Brueckner, E.~Marchi, M.~A. Nicolaou,
  B.~Schuller, and S.~Zafeiriou, ``Adieu features? end-to-end speech emotion
  recognition using a deep convolutional recurrent network,'' in \emph{IEEE
  International Conference on Acoustics, Speech and Signal Processing
  (ICASSP)}.\hskip 1em plus 0.5em minus 0.4em\relax IEEE, 2016, pp. 5200--5204.

\bibitem{hertel2016comparing}
L.~Hertel, H.~Phan, and A.~Mertins, ``Comparing time and frequency domain for
  audio event recognition using deep learning,'' in \emph{International Joint
  Conference on Neural Networks (IJCNN)}.\hskip 1em plus 0.5em minus
  0.4em\relax IEEE, 2016, pp. 3407--3411.

\bibitem{Hou2017}
Y.~Hou and S.~Li, ``Sound event detection in real life audio using multi-model
  system,'' DCASE2017 Challenge, Tech. Rep., September 2017.

\bibitem{cakir2016filterbank}
E.~Cakir, E.~C. Ozan, and T.~Virtanen, ``Filterbank learning for deep neural
  network based polyphonic sound event detection,'' in \emph{International
  Joint Conference on Neural Networks (IJCNN)}.\hskip 1em plus 0.5em minus
  0.4em\relax IEEE, 2016, pp. 3399--3406.

\bibitem{librosa}
\BIBentryALTinterwordspacing
B.~McFee, M.~McVicar, C.~Raffel, D.~Liang, O.~Nieto, E.~Battenberg, J.~Moore,
  D.~Ellis, R.~Yamamoto, R.~Bittner, D.~Repetto, P.~Viktorin, J.~F. Santos, and
  A.~Holovaty, ``librosa: 0.4.1,'' Oct. 2015. [Online]. Available:
  \url{http://dx.doi.org/10.5281/zenodo.32193}
\BIBentrySTDinterwordspacing

\bibitem{o1987speech}
D.~O'shaughnessy, \emph{Speech communication: human and machine}.\hskip 1em
  plus 0.5em minus 0.4em\relax Universities press, 1987.

\bibitem{cho2014properties}
K.~Cho, B.~Van~Merri{\"e}nboer, D.~Bahdanau, and Y.~Bengio, ``On the properties
  of neural machine translation: Encoder-decoder approaches,'' \emph{Eighth
  Workshop on Syntax, Semantics and Structure in Statistical Translation
  (SSST-8)}, 2014.

\bibitem{adamKeras}
D.~Kingma and J.~Ba, ``Adam: A method for stochastic optimization,'' in
  \emph{arXiv:1412.6980 [cs.LG]}, 2014.

\bibitem{batchNorm}
S.~Ioffe and C.~Szegedy, ``Batch normalization: Accelerating deep network
  training by reducing internal covariate shift,'' \emph{CoRR}, vol.
  abs/1502.03167, 2015.

\bibitem{Dropout}
N.~Srivastava, G.~Hinton, A.~Krizhevsky, I.~Sutskever, and R.~Salakhutdinov,
  ``Dropout: A simple way to prevent neural networks from overfitting,'' in
  \emph{Journal of Machine Learning Research (JMLR)}, 2014.

\bibitem{chollet2015keras}
F.~Chollet, ``Keras,'' \url{github.com/fchollet/keras}, 2015.

\bibitem{mesaros2016metrics}
A.~Mesaros, T.~Heittola, and T.~Virtanen, ``Metrics for polyphonic sound event
  detection,'' \emph{Applied Sciences}, vol.~6, no.~6, p. 162, 2016.

\end{thebibliography}

\end{document}